\documentclass[aps, prd, twocolumn, superscriptaddress]{revtex4-1}
\usepackage{graphicx}
\usepackage{amssymb}
\usepackage{amsmath}
\usepackage{booktabs}
\usepackage[colorlinks=True,linkcolor=blue,anchorcolor=blue,citecolor=blue,CJKbookmarks=true]{hyperref}
\usepackage{bbm}
\usepackage{bm}

\usepackage{ulem}

\begin{document}
\title{Nonadiabatic quantum kinetic equations and Dirac-Heisenberg-Wigner formalism for Schwinger pair production in time-varying electric fields with multiple components}

\author{Z. L. Li}
\email{zlli@cumtb.edu.cn}
\affiliation{School of Science, China University of Mining and Technology, Beijing 100083, China}
\affiliation{State Key Laboratory for Tunnel Engineering, China University of Mining and Technology, Beijing 100083, China}

\author{R. Z. Jiang}
\affiliation{State Key Laboratory for Tunnel Engineering, China University of Mining and Technology, Beijing 100083, China}

\author{Y. J. Li}
\email{lyj@aphy.iphy.ac.cn}
\affiliation{School of Science, China University of Mining and Technology, Beijing 100083, China}
\affiliation{State Key Laboratory for Tunnel Engineering, China University of Mining and Technology, Beijing 100083, China}

\date{\today}

\begin{abstract}
The nonadiabatic quantum kinetic equations and Dirac-Heisenberg-Wigner formalism for Schwinger pair production in a spatially uniform and time-varying electric field with multiple components are derived and proven to be equivalent. The relation between nonadiabatic and adiabatic quantum kinetic equations is also established. By analyzing the time evolution of the distribution functions of particles created in a circularly polarized Gaussian pulse field with a subcycle structure, it is found that the nonadiabatic and adiabatic distribution functions are the same after the field, with a sufficient number of oscillation cycles, fades away. However, during the presence of the field, the two distribution functions typically differ. Nonetheless, the time evolution characteristics of the nonadiabatic and adiabatic momentum distributions are similar. For instance, the number of spirals is one less than the number of photons absorbed in both cases. Furthermore, for a rapidly oscillating electric field, the nonadiabatic quantum kinetic approaches may provide a more meaningful description of pair production at intermediate times. These findings deepen our understanding of the nonadiabatic quantum kinetic approaches and their application in pair production.
\end{abstract}

\maketitle

\section{INTRODUCTION}

Quantum vacuum in the presence of an extremely strong electromagnetic field will become unstable and decay into electron-positron pairs \cite{Sauter1931,Heisenberg1936, Schwinger1951}. This process is one of the most fascinating phenomena predicted by quantum electrodynamics (QED), which is also called Schwinger pair production or Schwinger effect due to the pioneering work of Schwinger in 1951. While the Schwinger pair production was theoretically proposed decades ago, its direct experimental observation remains unrealized because of sufficiently strong external fields required. The critical electric field strength required for this process, known as the Schwinger limit, is $E_{\rm{cr}}=m^2/e\approx 1.3\times10^{18}\rm{V}/\rm{m}$, where $m$ is the electron mass and $e$ is the magnitude of the electron charge. Note that the natural units $\hbar=c=1$ are used throughout this paper. Achieving such extreme field strengths is beyond the capabilities of conventional laboratory setups. However, remarkable advancements in high-intensity laser technology and laser facilities \cite{Mourou2014,ELI} are expected to narrow the gap, bringing the experimental observation of this process tantalizingly close. This positions the Schwinger pair production as one of the focal points of strong-field physics \cite{Xie2017,Fedotov2023}.

Besides experimental progress, extensive theoretical studies have also been conducted on Schwinger pair production. They are mainly focus on exploring novel features in pair production \cite{HebenstreitPRL2009,Hebenstreit2011,Kohlfurst2014,LiPRD2015,Li2017} and lowering the threshold of pair production (or enhancing the yield of particle pairs) \cite{Schutzhold2008,Bell2008,Piazza2009,Bulanov2010,
LiPRD2014,Titov2012,Olugh2019}. Since the Schwinger pair production is a nonperturbative process, the frequently-used perturbative expansion in QED is not applicable and the nonperturbative methods are needed. These methods encompass: semiclassical approximations including the Wenzel-Kramers-Brillouin-like approximation \cite{Dumlu2010,Li2014-1,Oertel2019,Taya2021,Kohlfurst2022} and worldline instanton technique \cite{Affleck1982,Kim2002,Dunne2005,Dumlu2011,Schneider2018}, quantum kinetic approaches involving the quantum Vlasov equations (QVEs) \cite{Kluger1998,Schmidt1998,HebenstreitPRL2009,Kohlfurst2014} and Dirac-Heisenberg-Wigner (DHW) formalism \cite{Bialynicki-Birula1991,Hebenstreit2010,Hebenstreit2011,Li2017,Kohlfurst2020}, and numerical methods incorporating the computational quantum field theory \cite{Cheng2010,Jiang2012,Su2012,Su2019}. The relations between some of these methods are also established \cite{Dumlu2009,Hebenstreit2010,LiPRD2019,Li2021}, which can help us to verify and comprehend the computational results from different perspectives. As one of the quantum kinetic approaches, the QVEs have two advantages: first, they provide the time evolution of momentum distributions of created particles, enhancing our understanding of the pair production process; and second, they are a powerful tool for studying enhanced pair production through the optimization of field parameters \cite{KohlfurstPRD2013,HebenstreitPRD2014}. However, the QVEs are limited to studying pair production in spatially homogeneous and time-dependent electric fields. In particular, the QVEs in spinor QED can only address pair production in fields with a single component. Generally, to study pair production in multi-component fields, the DHW formalism is required, as it can handle pair production in both spatially and temporally varying fields. Recently, the QVEs have been generalized to quantum kinetic equations (QKEs) \cite{AleksandrovPRR2024}, which can study pair production in multi-component fields and have been proven to be equivalent to the DHW formalism for pair production in spatially uniform electric fields.

In addition, another version of the QVEs, referred to as nonadiabatic QVEs (NAQVEs), is proposed in \cite{Kim2011} based on the expansion of the Klein-Gordon field using plane wave solutions of free particles, and its connection with the QVEs is established in \cite{Huet2014}. Both of these findings are further generalized from scalar QED to spinor QED in \cite{LiPRD2024}. The NAQVEs provide us with another approach to investigate pair production in spatially uniform and time-varying electric fields, deepening the understanding of the results obtained from the QVEs through comparison with those computed using the NAQVEs. However, although the NAQVEs in scalar QED can be applied to study nonadiabatic pair production in multi-component electric fields, the NAQVEs in spinor QED are restricted to electric fields with a single component. And due to the absence of the nonadiabatic DHW formalism, some significant characteristics in nonadiabatic pair production, such as the effect of electric field polarization, the spiral structures, and the helicity signatures, can not be considered. To solve this problem, we will derive the nonadiabatic QKEs (NAQKEs) and the nonadiabatic DHW (NADHW) formalism in this paper and then demonstrate their equivalence. The relation between the NAQKEs and QKEs will also be established. As an example, we investigate the time evolution of nonadiabatic and adiabatic distribution functions of particles created in a circularly polarized Gaussian pulse electric field, which can reveal some important features of pair production at intermediate times.

The structure of this paper is organized as follows: In Sec. \ref{sec:two}, we derive the NAQKEs and NADHW formalism for Schwinger pair production in multi-component electric fields. In Sec. \ref{sec:three} the equivalence between the NAQKEs and NADHW formalism is proved. The relation between the NAQKEs and QKEs is established. It is further shown that the relation can recover the one between NAQVEs and QVEs for a linearly polarized electric field. In Sec. \ref{sec:four} the nonadiabatic pair production in a circularly polarized electric field with a subcycle structure is investigated by computing the time evolution of the nonadiabatic distribution functions and comparing it with that in the adiabatic case. Section \ref{sec:five} is a summary and discussion. For completeness, the nonadiabatic Feshbach-Villars-Heisenberg-Wigner (NAFVHW) formalism is derived, and its equivalence to the NAQVEs in scalar QED is demonstrated in Appendix \ref{appa}.

\section{Derivation of the nonadiabatic QKEs and DHW formalism}
\label{sec:two}

Our starting point is the Dirac equation for a spatially homogeneous and time-dependent electric field in the temporal gauge $A_\mu(\mathbf{x},t)=(0,-\mathbf{A}(t))$,
\begin{equation}\label{eqn:DiracEquation10}
\big\{i\gamma^{0}\partial_t+i\bm{\gamma}\cdot[\bm{\nabla}-i q \mathbf{A}(t)]-m\big\}\Psi(\mathbf{x},t)=0,
\end{equation}
where $\partial_t$ represents the partial derivative with respect to time, $\mathbf{A}(t)=(A_x(t), A_y(t), A_z(t))$ is the vector potential of the electric field, $q$ and $m$ are the charge and mass of the particle, respectively. $\gamma^0$ and $\bm{\gamma}$ are Dirac matrices.
The Fourier transform of the Dirac field has the form
\begin{equation}\label{eqn:FourierDeco0}
\Psi(\mathbf{x},t)=\int\frac{d^3k}{(2\pi)^3}\Psi_\mathbf{k}(t)e^{i\mathbf{k}\cdot\mathbf{x}},
\end{equation}
where the Fourier modes $\Psi_\mathbf{k}(t)$ satisfies
\begin{equation}\label{eqn:DiracEquation11}
\big[i\gamma^{0}\partial_t-\bm{\gamma}\cdot \mathbf{p}(t)-m \big]\Psi_\mathbf{k}(t)=0.
\end{equation}
Here $\mathbf{k}$ is the canonical momentum and $\mathbf{p}(t)=\mathbf{k}-q\mathbf{A}(t)$ is the kinetic momentum.
Equation (\ref{eqn:DiracEquation11}) can also be written as
\begin{equation}\label{eqn:DiracEquation12}
i\frac{\partial}{\partial t}\Psi_\mathbf{k}(t)=H_\mathbf{k}(t)\Psi_\mathbf{k}(t),
\end{equation}
where $H_\mathbf{k}(t)=\gamma^{0}\bm{\gamma}\cdot \mathbf{p}(t)+\gamma^{0}m$ is the Hamiltonian in momentum space.

In the absence of an external electric field, Eq. (\ref{eqn:DiracEquation10}) has the plane wave solutions with positive and negative energy states
\begin{equation}\label{eqn:u1}
u_{\mathbf{k},1}(t)=\left(
  \begin{array}{c}
   \omega_\mathbf{k}(t_0)+m \\
   0 \\
   p_z(t_0) \\
   p_x(t_0)+ip_y(t_0) \\
  \end{array}
\right)\chi^{+}_\mathbf{k}(t),
\end{equation}
\begin{equation}\label{eqn:u2}
u_{\mathbf{k},2}(t)=\left(
  \begin{array}{c}
    0 \\
    \omega_\mathbf{k}(t_0)+m \\
    p_x(t_0)-ip_y(t_0) \\
    -p_z(t_0) \\
  \end{array}
\right)\chi^{+}_\mathbf{k}(t),
\end{equation}
\begin{equation}\label{eqn:v1}
v_{-\mathbf{k},1}(t)=\left(
  \begin{array}{c}
   -p_z(t_0) \\
   -p_x(t_0)-ip_y(t_0) \\
   \omega_\mathbf{k}(t_0)+m \\
   0 \\
  \end{array}
\right)\chi^{-}_\mathbf{k}(t),\;\,
\end{equation}
\begin{equation}\label{eqn:v2}
v_{-\mathbf{k},2}(t)=\left(
  \begin{array}{c}
    -p_x(t_0)+ip_y(t_0) \\
    p_z(t_0) \\
    0 \\
    \omega_\mathbf{k}(t_0)+m \\
  \end{array}
\right)\chi^{-}_\mathbf{k}(t),
\end{equation}
where $t_0$ is the initial time, $\omega_\mathbf{k}(t_0)=[\mathbf{p}^2(t_0)+m^2]^{1/2}$, $\mathbf{p}(t_0)=\mathbf{k}-q\mathbf{A}(t_0)$, and
\begin{equation}\label{eqn:chit0}
\chi^{\pm}_\mathbf{k}(t)=\frac{e^{\mp i\omega_\mathbf{k}(t_0)(t-t_0)}}{\sqrt{2\omega_\mathbf{k}(t_0)[\omega
_\mathbf{k}(t_0)+m]}}.
\end{equation}
Furthermore, the orthonormal relations of positive and negative energy states are
\begin{equation}\label{eqn:Orthonormal1}
\begin{split}
u^\dagger_{\mathbf{k},s}(t)u_{\mathbf{k},s'}(t)=\delta_{ss'},\;\; v^\dagger_{-\mathbf{k},s}(t)v_{-\mathbf{k},s'}(t)=\delta_{ss'},\\
u^\dagger_{\mathbf{k},s}(t)v_{-\mathbf{k},s'}(t)=0, \;\; v^\dagger_{-\mathbf{k},s}(t)u_{\mathbf{k},s'}(t)=0,
\end{split}
\end{equation}
where $s, s'=\{1,2\}$ denote different spin states.

\subsection{Nonadiabatic QKEs}\label{sec:two-A}

The Dirac field operator can be expanded in terms of the plane wave solutions:
\begin{equation}\label{eqn:FieldOperator1}
\begin{split}
\hat{\Psi}(\mathbf{x},t)=\!\!\int\!\frac{d^3k}{(2\pi)^3}
\sum_{s=1}^2\big[&\,\hat{b}_{\mathbf{k},s}(t)
u_{\mathbf{k},s}(t)\\
&+\hat{d}_{-\mathbf{k},s}^{\,\dagger}(t)v_{-\mathbf{k},s}(t)\big]
e^{i\mathbf{k}\cdot\mathbf{x}},
\end{split}
\end{equation}
where $\hat{b}_{\mathbf{k},s}(t)$ and $\hat{d}_{-\mathbf{k},s}^{\dagger}(t)$ are the time-dependent annihilation and creation operators corresponding to the field-free vacuum state $|\mathrm{vac}\rangle_F$ and satisfy the usual equal-time anticommutation relations.

Since the Dirac field operator also satisfies Eq. (\ref{eqn:DiracEquation10}), by inserting Eq. (\ref{eqn:FieldOperator1}) into Eq. (\ref{eqn:DiracEquation10}) and using Eq. (\ref{eqn:Orthonormal1}), we can obtain
\begin{equation}\label{eqn:EvolutionOperator1}
\begin{split}
\dot{\hat{b}}_{\mathbf{k},s}(t)=-i\sum_{s''}[\hat{b}_{\mathbf{k},s''}(t)
\lambda_{1,ss''}+\hat{d}^\dagger_{-\mathbf{k},s''}(t)\lambda_{2,ss''}],\quad \\
\dot{\hat{d}}^\dagger_{-\mathbf{k},s}(t)=-i\sum_{s''}[\hat{b}_{\mathbf{k},s''}(t)
\lambda_{3,ss''}+\hat{d}^\dagger_{-\mathbf{k},s''}(t)\lambda_{4,ss''}],\quad
\end{split}
\end{equation}
where
\begin{equation}\label{eqn:Coefficient1}
\begin{split}
\lambda_{1,ss''}&=u_{\mathbf{k},s}^\dagger(t)H_\mathbf{k}(t)
u_{\mathbf{k},s''}(t)=\mu_1\delta_{ss''},\\
\lambda_{2,ss''}&=u_{\mathbf{k},s}^\dagger(t)H_\mathbf{k}(t)
v_{-\mathbf{k},s''}(t)=(\bm{\mu}_2\cdot\bm{\sigma})_{ss''},\\
\lambda_{3,ss''}&=v_{-\mathbf{k},s}^\dagger(t)H_\mathbf{k}(t)
u_{\mathbf{q},s''}(t)=(\bm{\mu}_2\cdot\bm{\sigma})_{ss''},\\
\lambda_{4,ss''}&=v_{-\mathbf{k},s}^\dagger(t)H_\mathbf{k}(t)
v_{-\mathbf{k},s''}(t)=-\mu_1\delta_{ss''},
\end{split}
\end{equation}
with
\begin{equation}\label{eqn:Coefficient2}
\begin{split}
\mu_1&=\frac{m^2+\mathbf{p}(t)\cdot\mathbf{p}(t_0)}{\omega_\mathbf{k}(t_0)}, \\
\bm{\mu}_2&=\mathbf{p}(t)-\frac{m^2+\mathbf{p}(t)\cdot\mathbf{p}(t_0)
+\omega_\mathbf{k}(t_0)m}{\omega_\mathbf{k}(t_0)[\omega_\mathbf{k}(t_0)+m]}
\mathbf{p}(t_0),
\end{split}
\end{equation}
and the Pauli matrices $\bm{\sigma}$.

The distribution function of particles created from the field-free vacuum $|\mathrm{vac}\rangle_F$ by an electric field is the diagonal components of the particle correlator
\begin{equation}\label{eqn:Correlator1}
f_{\mathbf{k},ss'}(t)=\langle \mathrm{vac}|\hat{b}^{\,\dagger}_{\mathbf{k},s'}(t)
\hat{b}_{\mathbf{k},s}(t)|\mathrm{vac}\rangle,
\end{equation}
where $|\mathrm{vac}\rangle$ is the vacuum state corresponding to the time-independent creation and annihilation operators used to expand the Dirac field operator.
The antiparticle correlator and two anomalous correlators are defined as
\begin{equation}\label{eqn:Correlator2}
\begin{split}
g_{\mathbf{k},ss'}(t)=\langle \mathrm{vac}|\hat{d}_{\mathbf{k},s'}(t)
\hat{d}^{\,\dagger}_{\mathbf{k},s}(t)|\mathrm{vac}\rangle,\\
y^+_{\mathbf{k},ss'}(t)=\langle \mathrm{vac}|\hat{b}^{\,\dagger}_{\mathbf{k},s'}(t)
\hat{d}^{\,\dagger}_{\mathbf{k},s}(t)|\mathrm{vac}\rangle,\\
y^-_{\mathbf{k},ss'}(t)=\langle \mathrm{vac}|\hat{d}_{\mathbf{k},s'}(t)
\hat{b}_{\mathbf{k},s}(t)|\mathrm{vac}\rangle.
\end{split}
\end{equation}
Differentiating the correlators above with respect to time and using Eqs. (\ref{eqn:EvolutionOperator1})-(\ref{eqn:Coefficient2}), we can obtain the matrix differential equaitons
\begin{equation}\label{eqn:MDE1}
\begin{split}
\dot{f}&=i(f\lambda_1-\lambda_1f-\lambda_2y^++y^-\lambda_2),\\
\dot{g}&=i(-g\lambda_1+\lambda_1g+y^+\lambda_2-\lambda_2y^-),\\
\dot{y}^+&=i(y^+\lambda_1+\lambda_1y^+-\lambda_2f+g\lambda_2),\\
\dot{y}^-&=i(-y^-\lambda_1-\lambda_1y^-+f\lambda_2-\lambda_2g),
\end{split}
\end{equation}
where $\lambda_1=\mu_1\mathbbm{1}$ and $\lambda_2=\bm{\mu}_2\cdot\bm{\sigma}$. By introducing two hermitian matrices
\begin{equation}\label{eqn:TwoHerM}
\begin{split}
h^+&=\frac{1}{2}(y^++y^-),\\
h^-&=\frac{i}{2}(y^+-y^-),
\end{split}
\end{equation}
and performing a spin decomposition of $\{f, g, h^\pm\}$, i.e., $f=f_0+\mathbf{f}\cdot\bm{\sigma}$, $g=g_0+\mathbf{g}\cdot\bm{\sigma}$, and $h^\pm=h^\pm_0+\mathbf{h}^\pm\cdot\bm{\sigma}$, Eqs. (\ref{eqn:MDE1}) become
\begin{eqnarray}
\dot{f}_0&=&-2\bm{\mu}_2\cdot\mathbf{h}^-,
\label{eqn:QKE-1}\\
\dot{\mathbf{f}}&=&2(\bm{\mu}_2\times\mathbf{h}^+)-2\bm{\mu}_2h^-_0,
\label{eqn:QKE-2}\\
\dot{g}_0&=&2\bm{\mu}_2\cdot\mathbf{h}^-,
\label{eqn:QKE-3}\\
\dot{\mathbf{g}}&=&2(\bm{\mu}_2\times\mathbf{h}^+)+2\bm{\mu}_2h^-_0,
\label{eqn:QKE-4}\\
\dot{h}^+_0&=&2\mu_1h^-_0,
\label{eqn:QKE-5}\\
\dot{\mathbf{h}}^+&=&2\mu_1\mathbf{h}^-+\bm{\mu}_2\times(\mathbf{f}
+\mathbf{g}),\label{eqn:QKE-6}\\
\dot{h}^-_0&=&-2\mu_1h^+_0+\bm{\mu}_2\cdot(\mathbf{f}-\mathbf{g}),
\label{eqn:QKE-7}\\
\dot{\mathbf{h}}^-&=&-2\mu_1\mathbf{h}^++\bm{\mu}_2(f_0-g_0).
\label{eqn:QKE-8}
\end{eqnarray}
The initial conditions are
\begin{equation}
\begin{split}
g_0(t_0)=1,\, f_0(t_0)=h^\pm_0(t_0)=0, \, \\ \mathbf{f}(t_0)=\mathbf{g}(t_0)=\mathbf{h}^\pm(t_0)=\mathbf{0}.
\end{split}
\end{equation}
According to the initial conditions and Eqs. (\ref{eqn:QKE-1}) and (\ref{eqn:QKE-3}), it finds that $g_0=1-f_0$. Furthermore, from Eqs. (\ref{eqn:QKE-2}), (\ref{eqn:QKE-4}), (\ref{eqn:QKE-5}), and (\ref{eqn:QKE-7}), it is easy to find that $h^\pm_0$ and $\mathbf{f}-\mathbf{g}$ form a closed homogeneous subsystem and $4(h^+_0)^2+4(h^-_0)^2+(\mathbf{f}-\mathbf{g})^2=0$. So we have $h^\pm_0=0$ and $\mathbf{f}=\mathbf{g}$. Finally, Eqs.(\ref{eqn:QKE-1})-(\ref{eqn:QKE-8}) are reduced to
\begin{eqnarray}
\dot{f}_0&=&-2\bm{\mu}_2\cdot\mathbf{h}^-,\label{eqn:QKE-11}\\
\dot{\mathbf{f}}&=&2(\bm{\mu}_2\times\mathbf{h}^+),\label{eqn:QKE-12}\\
\dot{\mathbf{h}}^+&=&2\mu_1\mathbf{h}^-+2(\bm{\mu}_2\times\mathbf{f}),\label{eqn:QKE-13}\\
\dot{\mathbf{h}}^-&=&-2\mu_1\mathbf{h}^++\bm{\mu}_2(2f_0-1),\label{eqn:QKE-14}
\end{eqnarray}
with initial conditions: $f_0(t_0)=0$, $\mathbf{f}(t_0)=\mathbf{h}^\pm(t_0)=\mathbf{0}$. These equations are the NAQKEs for pair production in three-component electric fields.

In the following we will show that the NAQKEs can recover the NAQVE  for a linearly polarized electric field.

In general, for an arbitrarily polarized electric field, assuming the unit vector of the electric field is $\mathbf{n}(t)$, the electric field can be expressed as $\mathbf{E}(t)=|\mathbf{E}(t)|\mathbf{n}(t)$. Similarly, the factor $\bm{\mu}_2$ in Eqs. (\ref{eqn:QKE-11})-(\ref{eqn:QKE-14}) can also be written as $\bm{\mu}_2(t)=|\bm{\mu}_2(t)|\mathbf{e}(t)$ using its unit vector $\mathbf{e}(t)$, where $|\bm{\mu}_2(t)|=\{\omega^2_\mathbf{k}(t)-[\frac{m^2+\mathbf{p}(t)\cdot
\mathbf{p}(t_0)}{\omega_\mathbf{k}(t_0)}]^2\}^{1/2}$ and $\omega_\mathbf{k}(t)=[\mathbf{p}^2(t)+m^2]^{1/2}$. However, for a linearly polarized electric field, the unit vector of the electric field does not change over time, i.e., $\dot{\mathbf{n}}(t)=0$. The unit vector $\mathbf{e}(t)$ does not change over time either. In this case, the electric field has the form $\mathbf{E}=E(t)\mathbf{n}$, and the factor becomes $\bm{\mu_2}(t)=\mu_2(t)\mathbf{e}$, where
\begin{equation}\label{eqn:mu2}
\mu_2(t)=\frac{\epsilon_\perp[\mathbf{p}(t)\cdot{\mathbf{n}}
-\mathbf{p}(t_0)\cdot{\mathbf{n}}]}{\omega_\mathbf{k}(t_0)},
\end{equation}
$\epsilon_\perp=(\mathbf{p}^2_\perp+m^2)^{1/2}$, $\mathbf{p}_\perp$ is the momentum perpendicular to $\mathbf{n}$, and $\mathbf{e}$ is the unit vector that forms an acute angle with $\mathbf{n}$.
Decomposing $\mathbf{f}$ and $\mathbf{h}^\pm$ in Eqs. (\ref{eqn:QKE-11})-(\ref{eqn:QKE-14}) along the directions parallel and perpendicular to $\mathbf{e}$ as $\mathbf{f}=f_\parallel\mathbf{e}+\mathbf{f}_\perp$ and $\mathbf{h}^\pm=h^\pm_\parallel\mathbf{e}+\mathbf{h}^\pm_\perp$, we get
\begin{eqnarray}
\dot{f}_0&=&-2\mu_2h^-_\parallel,\label{eqn:QVE-1}\\
\dot{f}_\parallel&=&0,\label{eqn:QVE-2}\\
\dot{h}^+_\parallel&=&2\mu_1h^-_\parallel,\label{eqn:QVE-3}\\
\dot{h}^-_\parallel&=&-2\mu_1h^+_\parallel+\mu_2(2f_0-1), \label{eqn:QVE-4}\\
\dot{\mathbf{f}}_\perp&=&2\mu_2(\mathbf{e}\times\mathbf{h}^+_\perp),
\label{eqn:QVE-5}\\
\dot{\mathbf{h}}^+_\perp&=&2\mu_1\mathbf{h}^-_\perp+2\mu_2(\mathbf{e}
\times\mathbf{f}_\perp),\label{eqn:QVE-6}\\
\dot{\mathbf{h}}^-_\perp&=&-2\mu_1\mathbf{h}^+_\perp.\label{eqn:QVE-7}
\end{eqnarray}
From Eqs. (\ref{eqn:QVE-5})-(\ref{eqn:QVE-7}), one can find that $\mathbf{f}^2_\perp+(\mathbf{h}^+_\perp)^2+(\mathbf{h}^-_\perp)^2
=\mathrm{Constant}$. According to the initial conditions, we have $f_\parallel=0, \mathbf{f}_\perp=\mathbf{h}^\pm_\perp=\mathbf{0}$. Eqs. (\ref{eqn:QVE-1})-(\ref{eqn:QVE-7}) become
\begin{equation}\label{eqn:QVE-11}
\begin{split}
\dot{f}_0&=-2\mu_2h^-_\parallel,\\
\dot{h}^+_\parallel&=2\mu_1h^-_\parallel,\\
\dot{h}^-_\parallel&=-2\mu_1h^+_\parallel+\mu_2(2f_0-1).
\end{split}
\end{equation}
The initial conditions are $f_0(t_0)=0$, $f_\parallel(t_0)=h^\pm_\parallel(t_0)=0$.
With the replacement, $f_0\rightarrow f^+_\mathbf{k}, h^+_\parallel\rightarrow g^+_\mathbf{k}/2, h^-_\parallel\rightarrow h^+_\mathbf{k}/2, \mu_1\rightarrow\Omega^+_\mathbf{k}, \mu_2\rightarrow -Q^+_\mathbf{k}/2$, one can see that Eqs. (\ref{eqn:QVE-11}) are exactly the form of ordinary differential equations of NAQVE, see Eqs. (25) in Ref. \cite{LiPRD2024}.

\subsection{Nonadiabatic DHW formalism}\label{sec:two-B}

The Wigner function is defined based on the equal-time commutator of two Dirac field operators in the Heisenberg picture:
\begin{eqnarray}\label{eqn:WignerFunction1}
\mathcal{W}(\mathbf{x},\mathbf{p},t)=-\tfrac{1}{2}\int &&d^3r\,e^{-i\mathbf{p}\cdot\mathbf{r}}
e^{-iq\int_{-\frac{1}{2}}^{\frac{1}{2}}d\xi\mathbf{A}(\mathbf{x}
+\xi\mathbf{r},t)\cdot\mathbf{r}\,}\,\\
&&\times\langle\mathrm{vac}|[\hat{\Psi}(\mathbf{x}+\tfrac{\mathbf{r}}{2},t),
\hat{\bar{\Psi}}(\mathbf{x}-\tfrac{\mathbf{r}}{2},t)]|\mathrm{vac}\rangle. \nonumber
\end{eqnarray}
where $\mathbf{x}$ is the center-of-mass coordinate, $\mathbf{r}$ is the relative coordinate, $\mathbf{p}$ is the kinetic momentum, and $\hat{\bar{\Psi}}=\hat{\Psi}^\dagger\gamma^0$. Note that the second exponential function in the integrand is introduced to keep gauge invariance.

The Wigner function can be decomposed by a complete basis set
$\{\mathbbm{1},\gamma_5,\gamma^\mu,\gamma^\mu\gamma_5,\sigma^{\mu\nu}
=:\frac{i}{2}[\gamma^\mu,\gamma^\nu]\}$ as
\begin{equation}\label{eqn:WignerExpand1}
\mathcal{W}=\frac{1}{4}\left(\mathbbm{1}\mathbbm{s}+i\gamma_5\mathbbm{p}
+\gamma^\mu\mathbbm{v}_\mu+\gamma^\mu\gamma_5\mathbbm{a}_\mu
+\sigma^{\mu\nu}\mathbbm{t}_{\mu\nu}\right),
\end{equation}
where $\mathbbm{s}$, $\mathbbm{p}$, $\mathbbm{v}_\mu$, $\mathbbm{a}_\mu$, $\mathbbm{t}_{\mu\nu}$ are scalar, pseudoscalar, vector, axialvector and tensor, respectively. These sixteen irreducible DHW components depend on $\mathbf{x}$, $\mathbf{p}$, and $t$. Substituting the decomposition above into the equation of motion of the Wigner function, one can get sixteen partial differential equations. In particular, for a spatially homogeneous and time dependent electric field, these partial differential equations can be simplified to ten ordinary differential equations:
\begin{eqnarray}
\dot{\mathbbm{s}}&=&2\mathbf{p}\cdot\bm{\mathbbm{t}},\label{eqn:DHW-1}\\
\dot{\bm{\mathbbm{v}}}&=&-2\mathbf{p}\times\bm{\mathbbm{a}}
-2m\bm{\mathbbm{t}},\label{eqn:DHW-2}\\
\dot{\bm{\mathbbm{a}}}&=&-2\mathbf{p}\times\bm{\mathbbm{v}},
\label{eqn:DHW-3}\\
\dot{\bm{\mathbbm{t}}}&=&-2\mathbf{p}\mathbbm{s}+2m\bm{\mathbbm{v}},
\label{eqn:DHW-4}
\end{eqnarray}
where $\bm{\mathbbm{v}}$ and $\bm{\mathbbm{a}}$ are the spatial components of $\mathbbm{v}_\mu$ and $\mathbbm{a}_\mu$, $\bm{\mathbbm{t}}=2\mathbbm{t}^{i0}\mathbf{e}_i$, $\mathbf{e}_i$ is the unit vector in the $i$-direction. Notice that these ten DHW components depend only on $\mathbf{p}$ and $t$. The initial conditions are $\bm{\mathbbm{a}}(t_0)=\bm{\mathbbm{t}}(t_0)=\mathbf{0}$,
$\mathbbm{s}(t_0)=-2m/\omega_{\mathbf{p}(t_0)}$, and $\bm{\mathbbm{v}}(t_0)=-2\mathbf{p}(t_0)/\omega_{\mathbf{p}(t_0)},
$ where $\omega_{\mathbf{p}(t_0)}=[\mathbf{p}^2(t_0)+m^2]^{1/2}
=\omega_{\mathbf{k}}(t_0)$.

As can be seen from Eq. (\ref{eqn:WignerFunction1}), the Wigner function does not depend on the specific expansion form of the Dirac field operator. So the DHW components are independent of what basis the field operator uses to expand. However, the definition of particle number is related to the basis used to expand the field operator. Specifically, it is related to the vacuum state corresponding to the creation and annihilation operators in the expansion. Generally, the distribution function of particles created from the vacuum $|\mathrm{vac}\rangle_X$ is defined as
\begin{equation}\label{eqn:ParticleDefinition1}
f_{\mathbf{p}_X}(t)=\frac{\epsilon_{\mathbf{p}_X}(t)-
\epsilon^\mathrm{vac}_{\mathbf{p}_X}}{2\omega_{\mathbf{p}_X}},
\end{equation}
where $\epsilon_{\mathbf{p}_X}(t)=m\mathbbm{s}+\mathbf{p}_X
\cdot\bm{\mathbbm{v}}$ denotes the phase-space energy density of created particles, $\mathbf{p}_X$ is the kinetic momentum of the particle, $\epsilon^\mathrm{vac}_{\mathbf{p}_X}=m\mathbbm{s}^\mathrm{vac}
+\mathbf{p}_X\cdot\bm{\mathbbm{v}}^\mathrm{vac}$ is the phase-space energy density of the vacuum state $|\mathrm{vac}\rangle_X$,
\begin{equation}\label{eqn:VacuumStates1}
\mathbbm{s}^\mathrm{vac}=-\frac{2m}{\omega_{\mathbf{p}_X}},\quad \bm{\mathbbm{v}}^\mathrm{vac}=-\frac{2\mathbf{p}_X}{\omega_{\mathbf{p}_X}},
\end{equation}
and $\omega_{\mathbf{p}_X}=(\mathbf{p}^2_X+m^2)^{1/2}$ is the energy of the particle corresponding to the vacuum state. The definition in Eq. (\ref{eqn:ParticleDefinition1}) can be understood as follows. First, according to Eq. (\ref{eqn:VacuumStates1}), one can find that $\epsilon^\mathrm{vac}_{\mathbf{p}_X}=-2\omega_{\mathbf{p}_X}$. This result can be understood by Dirac sea picture. We know that the Dirac vacuum is a state in which the negative energy sea is filled with negative energy electrons, and the positive energy state is not occupied by electrons. Since each negative energy state is occupied by two negative energy electrons with opposite spins but the same energy $-\omega_{\mathbf{p}_X}$, the phase-space energy density of the vacuum state is $-2\omega_{\mathbf{p}_X}$. It is easy to see that the vacuum energy density is divergent. Secondly, the denominator in the definition (\ref{eqn:ParticleDefinition1}) represents the energy required to produce a pair of electron and positron with opposite momentum $\mathbf{p}_X$ and the same energy $\omega_{\mathbf{p}_X}$, so Eq. (\ref{eqn:ParticleDefinition1}) shows the total number of created electron-positron pairs with opposite momentum $\mathbf{p}_X$ and the same energy $\omega_{\mathbf{p}_X}$. Finally, it should be point out that the distribution function defined in the commonly used DHW formalism is based on the adiabatic vacuum state,  $|\mathrm{vac}\rangle_A$, which is related to the instantaneous eigenstate of the time-dependent Hamiltonian. In this case, the kinetic momentum $\mathbf{p}_X$ is $\mathbf{p}(t)$. If we choose the initial field-free vacuum state $|\mathrm{vac}\rangle_F$, the kinetic momentum $\mathbf{p}_X$ in the definition of the distribution function becomes $\mathbf{p}(t_0)$ and Eq. (\ref{eqn:ParticleDefinition1}) becomes
\begin{equation}\label{eqn:ParticleDefinition2}
f_{\mathbf{p}(t_0)}(t)=\frac{m\mathbbm{s}+\mathbf{p}(t_0)
\cdot\bm{\mathbbm{v}}}{2\omega_{\mathbf{p}(t_0)}}+1.
\end{equation}
This equation and Eqs. (\ref{eqn:DHW-1})-(\ref{eqn:DHW-4}) together constitute the NADHW formalism for Schwinger pair production in spatially homogeneous and time-dependent electric fields with three components.

In order to prove the equivalence of the NAQKEs and NADHW formalism in the next section, we give another version of the NADHW formalism by replacing $\mathbbm{s}$ and $\bm{\mathbbm{v}}$ in Eqs. (\ref{eqn:DHW-1})-(\ref{eqn:DHW-4}) with $2m[f_{\mathbf{p}(t_0)}(t)-1]/\omega_{\mathbf{p}(t_0)}-\mathbf{p}(t_0)
\cdot\bm{\mathfrak{v}}/m$ and $\bm{\mathfrak{v}}+2\mathbf{p}(t_0)[f_{\mathbf{p}(t_0)}(t)-1]
/\omega_{\mathbf{p}(t_0)}$,
\begin{eqnarray}
\dot{f}_0&=&\frac{1}{2\omega_0}[m(\mathbf{p}-\mathbf{p}_0)\cdot
\bm{\mathbbm{t}}-\mathbf{p}_0\cdot(\mathbf{p}\times\bm{\mathbbm{a}})],
\label{eqn:DHW-11}\\
\dot{\bm{\mathfrak{v}}}&=&-2\mathbf{p}\times\bm{\mathbbm{a}}
-2m\bm{\mathbbm{t}}-\frac{4\mathbf{p}_0}{\omega^2_0}
[m(\mathbf{p}-\mathbf{p}_0)\cdot\bm{\mathbbm{t}}\nonumber\\
&&-\mathbf{p}_0\cdot(\mathbf{p}\times\bm{\mathbbm{a}})],\label{eqn:DHW-12}\\
\dot{\bm{\mathbbm{a}}}&=&-2\mathbf{p}\times\bm{\mathfrak{v}}
-\frac{4}{\omega_0}(\mathbf{p}\times\mathbf{p}_0)(2f_0-1),
\label{eqn:DHW-13}\\
\dot{\bm{\mathbbm{t}}}&=&\frac{2}{m}(\mathbf{p}_0\cdot
\bm{\mathfrak{v}})\mathbf{p}+2m\bm{\mathfrak{v}}-\frac{4m}{\omega_0}
(\mathbf{p}-\mathbf{p}_0)(2f_0-1). \;
\label{eqn:DHW-14}
\end{eqnarray}
The initial conditions are $f_0=0$, $\bm{\mathfrak{v}}=\bm{\mathbbm{a}}=\bm{\mathbbm{t}}=\mathbf{0}$. Note that $ \mathbf{p}(t), \mathbf{p}(t_0), f_{\mathbf{p}(t_0)}(t), \omega_{\mathbf{p}(t_0)}$ are re-expressed as $\mathbf{p}, \mathbf{p}_0, 2f_0, \omega_0$ for convenience.

\section{Relating NAQKEs to NADHW formalism and QKEs}
\label{sec:three}

In this section, we will prove the equivalence of NAQKEs and NADHW formalism and establish the relation between the NAQKEs and QKEs.

\subsection{Equivalence of NAQKEs and NADHW formalism}\label{A}

First, we derive the transformation from Eqs. (\ref{eqn:DHW-1})-(\ref{eqn:DHW-4}) to Eqs. (\ref{eqn:QKE-11})-(\ref{eqn:QKE-14}).
Substituting Eq. (\ref{eqn:FieldOperator1}) into Eq. (\ref{eqn:WignerFunction1}) and using the definitions of correlators [Eqs. (\ref{eqn:Correlator1}), (\ref{eqn:Correlator2}), and (\ref{eqn:TwoHerM})] and their spin decomposition, we get
\begin{equation}\label{eqn:WignerFunction2}
\begin{split}
\hspace{-0.2cm}\mathcal{W}(\mathbf{p},t)=&\Sigma_{s}[f_0u_{\mathbf{k},s}
\bar{u}_{\mathbf{k},s}\!+\!(1\!-\!f_0)v_{-\mathbf{k},s}
\bar{v}_{-\mathbf{k},s}]\!-\!\frac{1}{2}\gamma^0\\
&+\Sigma_{ss'}\{(\mathbf{f}\cdot\bm{\sigma})_{ss'}[u_{\mathbf{k},s}
\bar{u}_{\mathbf{k},s'}+v_{-\mathbf{k},s}\bar{v}_{-\mathbf{k},s'}]\\
&+(\mathbf{h}^+\cdot\bm{\sigma})_{ss'}[u_{\mathbf{k},s}
\bar{v}_{-\mathbf{k},s'}+v_{-\mathbf{k},s}\bar{u}_{\mathbf{k},s'}]\\
&+i(\mathbf{h}^-\cdot\bm{\sigma})_{ss'}[u_{\mathbf{k},s}
\bar{v}_{-\mathbf{k},s'}-v_{-\mathbf{k},s}\bar{u}_{\mathbf{k},s'}]\},
\end{split}
\end{equation}
where $\mathbf{k}=\mathbf{p}(t)+q\mathbf{A}(t)$ is just the canonical momentum defined in Eq. (\ref{eqn:FourierDeco0}). Utilizing Eqs. (\ref{eqn:u1})-(\ref{eqn:chit0}) and
$\mathbbm{s}=\mathrm{Tr}(\mathcal{W}),\, \bm{\mathbbm{v}}=\mathrm{Tr}(\bm{\gamma}\mathcal{W}),\, \bm{\mathbbm{a}}=\mathrm{Tr}(\gamma_5\bm{\gamma}\mathcal{W}), \, \bm{\mathbbm{t}}=-\mathrm{Tr}(\sigma^{0i}\mathcal{W})$, we finally obtain the transformation
\begin{eqnarray}
\mathbbm{s}&=&\frac{4}{\omega_0}\Big[m\Big(f_0-\frac{1}{2}\Big)
-\mathbf{p}_0\cdot\mathbf{h}^+\Big],
\label{eqn:TF-1}\\
\bm{\mathbbm{v}}&=&\frac{4}{\omega_0}\Big[\omega_0\mathbf{h}^+
+\mathbf{p}_0\Big(f_0-\frac{1}{2}\Big)-\frac{\mathbf{p}_0(\mathbf{p}_0
\cdot\mathbf{h}^+)}{\omega_0+m}\Big],\label{eqn:TF-2}\\
\bm{\mathbbm{a}}&=&-\frac{4}{\omega_0}\Big[m\mathbf{f}
-\mathbf{p}_0\times\mathbf{h}^-+\frac{\mathbf{p}_0(\mathbf{p}_0
\cdot\mathbf{f})}{\omega_0+m}\Big],
\label{eqn:TF-3}\\
\bm{\mathbbm{t}}&=&-\frac{4}{\omega_0}\Big[m\mathbf{h}^-
-\mathbf{p}_0\times\mathbf{f}+\frac{\mathbf{p}_0(\mathbf{p}_0
\cdot\mathbf{h}^-)}{\omega_0+m}\Big].
\label{eqn:TF-4}
\end{eqnarray}
The inverse transformation is
\begin{eqnarray}
f_0&=&\frac{1}{4\omega_0}(m\mathbbm{s}+\mathbf{p}_0
\cdot\bm{\mathbbm{v}})+\frac{1}{2}, \label{eqn:TFI-1}\\
\mathbf{f}&=&-\frac{1}{4\omega_0}\Big[m\bm{\mathbbm{a}}
+\mathbf{p}_0\times\bm{\mathbbm{t}}+\frac{\mathbf{p}_0(\mathbf{p}_0
\cdot\bm{\mathbbm{a}})}{\omega_0+m}\Big],\label{eqn:TFI-2}\\
\mathbf{h}^+&=&-\frac{1}{4\omega_0}\Big[\mathbf{p}_0\mathbbm{s}
-\omega_0\bm{\mathbbm{v}}+\frac{\mathbf{p}_0(\mathbf{p}_0
\cdot\bm{\mathbbm{v}})}{\omega_0+m}\Big],\label{eqn:TFI-3}\\
\mathbf{h}^-&=&-\frac{1}{4\omega_0}\Big[m\bm{\mathbbm{t}}
+\mathbf{p}_0\times\bm{\mathbbm{a}}+\frac{\mathbf{p}_0(\mathbf{p}_0
\cdot\bm{\mathbbm{t}})}{\omega_0+m}\Big]. \label{eqn:TFI-4}
\end{eqnarray}

For another version of the NADHW formalism, the transformation from NADHW formalism [Eqs. (\ref{eqn:DHW-11})-(\ref{eqn:DHW-14})] to NAQVEs [Eqs. (\ref{eqn:QKE-11})-(\ref{eqn:QKE-14})] is
\begin{eqnarray}
f_0&=&f_0, \label{eqn:TF-21}\\
\bm{\mathfrak{v}}&=&\frac{4}{\omega_0}\Big[\omega_0\mathbf{h}^+
-\frac{\mathbf{p}_0(\mathbf{p}_0\cdot\mathbf{h}^+)}{\omega_0+m}\Big],
\label{eqn:TF-22}\\
\bm{\mathbbm{a}}&=&-\frac{4}{\omega_0}\Big[m\mathbf{f}
-\mathbf{p}_0\times\mathbf{h}^-+\frac{\mathbf{p}_0(\mathbf{p}_0
\cdot\mathbf{f})}{\omega_0+m}\Big],
\label{eqn:TF-23}\\
\bm{\mathbbm{t}}&=&-\frac{4}{\omega_0}\Big[m\mathbf{h}^-
-\mathbf{p}_0\times\mathbf{f}+\frac{\mathbf{p}_0(\mathbf{p}_0
\cdot\mathbf{h}^-)}{\omega_0+m}\Big].
\label{eqn:TF-24}
\end{eqnarray}
The inverse transformation is
\begin{eqnarray}
f_0&=&f_0, \label{eqn:TFI-21}\\
\mathbf{f}&=&-\frac{1}{4\omega_0}\Big[m\bm{\mathbbm{a}}
+\mathbf{p}_0\times\bm{\mathbbm{t}}+\frac{\mathbf{p}_0(\mathbf{p}_0
\cdot\bm{\mathbbm{a}})}{\omega_0+m}\Big], \label{eqn:TFI-22}\\
\mathbf{h}^+&=&\frac{1}{4m}\Big[m\bm{\mathfrak{v}}
+\frac{\mathbf{p}_0(\mathbf{p}_0\cdot\bm{\mathfrak{v}})}{\omega_0+m}
\Big],\label{eqn:TFI-23}\\
\mathbf{h}^-&=&-\frac{1}{4\omega_0}\Big[m\bm{\mathbbm{t}}
+\mathbf{p}_0\times\bm{\mathbbm{a}}+\frac{\mathbf{p}_0(\mathbf{p}_0
\cdot\bm{\mathbbm{t}})}{\omega_0+m}\Big]. \label{eqn:TFI-24}
\end{eqnarray}
As can be seen from the transformation, the NADHW formalism is equivalent to the NAQKEs.

\subsection{Relation between NAQKEs and QKEs}\label{B}

In this section, we will use DHW formalism [Eqs. (\ref{eqn:DHW-1})-(\ref{eqn:DHW-4})] as a bridge to establish the relation between NAQKEs and QKEs, because these equations are independent of the expansion form of the field operator and are related to both NAQKEs and QKEs.

From Eqs. (154)-(161) in  Ref. \cite{AleksandrovPRR2024}, we know that the transformation from the DHW formalism to QKEs is
\begin{eqnarray}
\mathbbm{s}&=&\frac{4}{\omega}\Big[m\Big(f-\frac{1}{2}\Big)
-\mathbf{p}\cdot\mathbf{u}\Big],
\label{eqn:TF-31}\\
\bm{\mathbbm{v}}&=&\frac{4}{\omega}\Big[\omega\mathbf{u}
+\mathbf{p}\Big(f-\frac{1}{2}\Big)-\frac{\mathbf{p}(\mathbf{p}
\cdot\mathbf{u})}{\omega+m}\Big],\label{eqn:TF-32}\\
\bm{\mathbbm{a}}&=&-\frac{4}{\omega}\Big[m\mathbf{f}
-\mathbf{p}\times\mathbf{v}+\frac{\mathbf{p}(\mathbf{p}
\cdot\mathbf{f})}{\omega+m}\Big],
\label{eqn:TF-33}\\
\bm{\mathbbm{t}}&=&-\frac{4}{\omega}\Big[m\mathbf{v}
-\mathbf{p}\times\mathbf{f}+\frac{\mathbf{p}(\mathbf{p}
\cdot\mathbf{v})}{\omega+m}\Big],
\label{eqn:TF-34}
\end{eqnarray}
where $f=f_{\mathbf{p}(t)}(t)$ is the distribution function of created particles from the adiabatic vacuum, $\omega=\omega_{\mathbf{p}(t)}=\omega_\mathbf{k}(t)$.
The inverse transformation is
\begin{eqnarray}
f&=&\frac{1}{4\omega}(m\mathbbm{s}+\mathbf{p}
\cdot\bm{\mathbbm{v}})+\frac{1}{2}, \label{eqn:TFI-31}\\
\mathbf{f}&=&-\frac{1}{4\omega}\Big[m\bm{\mathbbm{a}}
+\mathbf{p}\times\bm{\mathbbm{t}}+\frac{\mathbf{p}(\mathbf{p}
\cdot\bm{\mathbbm{a}})}{\omega+m}\Big],\label{eqn:TFI-32}\\
\mathbf{u}&=&-\frac{1}{4\omega}\Big[\mathbf{p}\mathbbm{s}
-\omega\bm{\mathbbm{v}}+\frac{\mathbf{p}(\mathbf{p}
\cdot\bm{\mathbbm{v}})}{\omega+m}\Big],\label{eqn:TFI-33}\\
\mathbf{v}&=&-\frac{1}{4\omega}\Big[m\bm{\mathbbm{t}}
+\mathbf{p}\times\bm{\mathbbm{a}}+\frac{\mathbf{p}(\mathbf{p}
\cdot\bm{\mathbbm{t}})}{\omega+m}\Big]. \label{eqn:TFI-34}
\end{eqnarray}
Therefore, the transformation from NAQKEs [QKEs] to QKEs [NAQKEs] can be obtained by substituting Eqs. (\ref{eqn:TF-31})-(\ref{eqn:TF-34}) [Eqs. (\ref{eqn:TF-1})-(\ref{eqn:TF-4})] into Eqs. (\ref{eqn:TFI-1})-(\ref{eqn:TFI-4}) [Eqs. (\ref{eqn:TFI-31})-(\ref{eqn:TFI-34})]. In particular, we get the relation between the distribution function $f_0$ and $f$ as
\begin{equation}\label{eqn:f0-f}
\begin{split}
f_0=&\frac{m^2+\mathbf{p}_0\cdot\mathbf{p}}{\omega_0\omega}
\Big(f-\frac{1}{2}\Big)+\frac{1}{2}\\
&+\frac{1}{\omega_0}\Big[\mathbf{p}_0
-\frac{m^2+\mathbf{p}_0\cdot\mathbf{p}+\omega m}{\omega(\omega+m)}\mathbf{p}\Big]\cdot\mathbf{u}.
\end{split}
\end{equation}
The inverse transformation is
\begin{equation}\label{eqn:f-f0}
\begin{split}
f=&\frac{m^2+\mathbf{p}_0\cdot\mathbf{p}}{\omega_0\omega}
\Big(f_0-\frac{1}{2}\Big)+\frac{1}{2}\\
&+\frac{1}{\omega}\Big[\mathbf{p}
-\frac{m^2+\mathbf{p}_0\cdot\mathbf{p}+\omega_0 m}{\omega_0(\omega_0+m)}\mathbf{p}_0\Big]\cdot\mathbf{h}^+.
\end{split}
\end{equation}

For a linearly polarized electric field, Eq. (\ref{eqn:f0-f}) can return to the relation between NAQVE and QVE [Eqs. (37) and (38)] founded in Ref. \cite{LiPRD2024}.

Supposing $\mathbf{n}$ is the unit vector of the electric field and $\mathbf{e}$ is the time-independent unit vector of $\bm{\mu}_2(t)$ in Ref. \cite{AleksandrovPRR2024}, then Eq. (\ref{eqn:f0-f}) becomes
\begin{equation}\label{eqn:NAQVE-QVE}
\begin{split}
f_0=&\frac{\epsilon_\perp^2+(\mathbf{p}_0\cdot\mathbf{n})
(\mathbf{p}\cdot\mathbf{n})}{\omega_0\omega}
\Big(f-\frac{1}{2}\Big)+\frac{1}{2}\\
&+\frac{\epsilon_\perp[(\mathbf{p}\cdot\mathbf{n})-(\mathbf{p}_0
\cdot\mathbf{n})]}{\omega\omega_0}
(\mathbf{e}\cdot\mathbf{u}),
\end{split}
\end{equation}
where $(\mathbf{e}\cdot\mathbf{u})$ is the $u_\parallel$ defined in Eqs. (109) in Ref. \cite{AleksandrovPRR2024}. Furthermore,  $u_\parallel=-u(\mathbf{k},t)/2$ and $u(\mathbf{k},t)$ is defined in Eq. (3) in Ref. \cite{LiPRD2014}. So we can see that the above equation shows exactly the relation between NAQVE and QVE. In this case, the inverse relation between them is
\begin{equation}\label{eqn:QVE-NAQVE}
\begin{split}
f=&\frac{\epsilon_\perp^2+(\mathbf{p}_0\cdot\mathbf{n})
(\mathbf{p}\cdot\mathbf{n})}{\omega_0\omega}
\Big(f_0-\frac{1}{2}\Big)+\frac{1}{2}\\
&+\frac{\epsilon_\perp[(\mathbf{p}\cdot\mathbf{n})-(\mathbf{p}_0
\cdot\mathbf{n})]}{\omega\omega_0}
h^+_\parallel.
\end{split}
\end{equation}
where $h^+_\parallel$ is defined in Eq. (\ref{eqn:QVE-11}) and equals $g^+_\mathbf{k}/2$ used in Eq. (25) in Ref. \cite{LiPRD2024}.

In addition, we also derive the nonadiabatic FVHW formalism and prove that it is equivalent to the NAQVEs in scalar QED, see Appendix \ref{appa}. Since both NAQVEs and QVEs in scalar QED apply to pair production in multi-component electric fields, the relation between them derived in Ref. \cite{LiPRD2024} is still valid for pair production in multi-component electric fields.

\section{Time evolution of distribution functions of created particles}
\label{sec:four}

From Eqs. (\ref{eqn:f0-f}) and (\ref{eqn:f-f0}), one can see that the distribution function of particles created from field-free vacuum (nonadiabatic distribution function) is generally not equal to that created from adiabatic vacuum (adiabatic distribution function) unless $\mathbf{A}(t)=\mathbf{A}(t_0)$ is satisfied. For instance, the two distribution functions are different for a single Sauter electric field even after the field vanishes, because the vector potential $\mathbf{A}(t_f)\neq\mathbf{A}(t_0)$, where $t_f$ is the time after the electric field vanishes. However, we will show that the nonadiabatic and adiabatic distribution functions at time  $t_f$ are almost identical for a Gaussian envelope pulse with a subcycle structure.

\begin{figure}[!ht]
\centering
\includegraphics[width=0.4\textwidth]{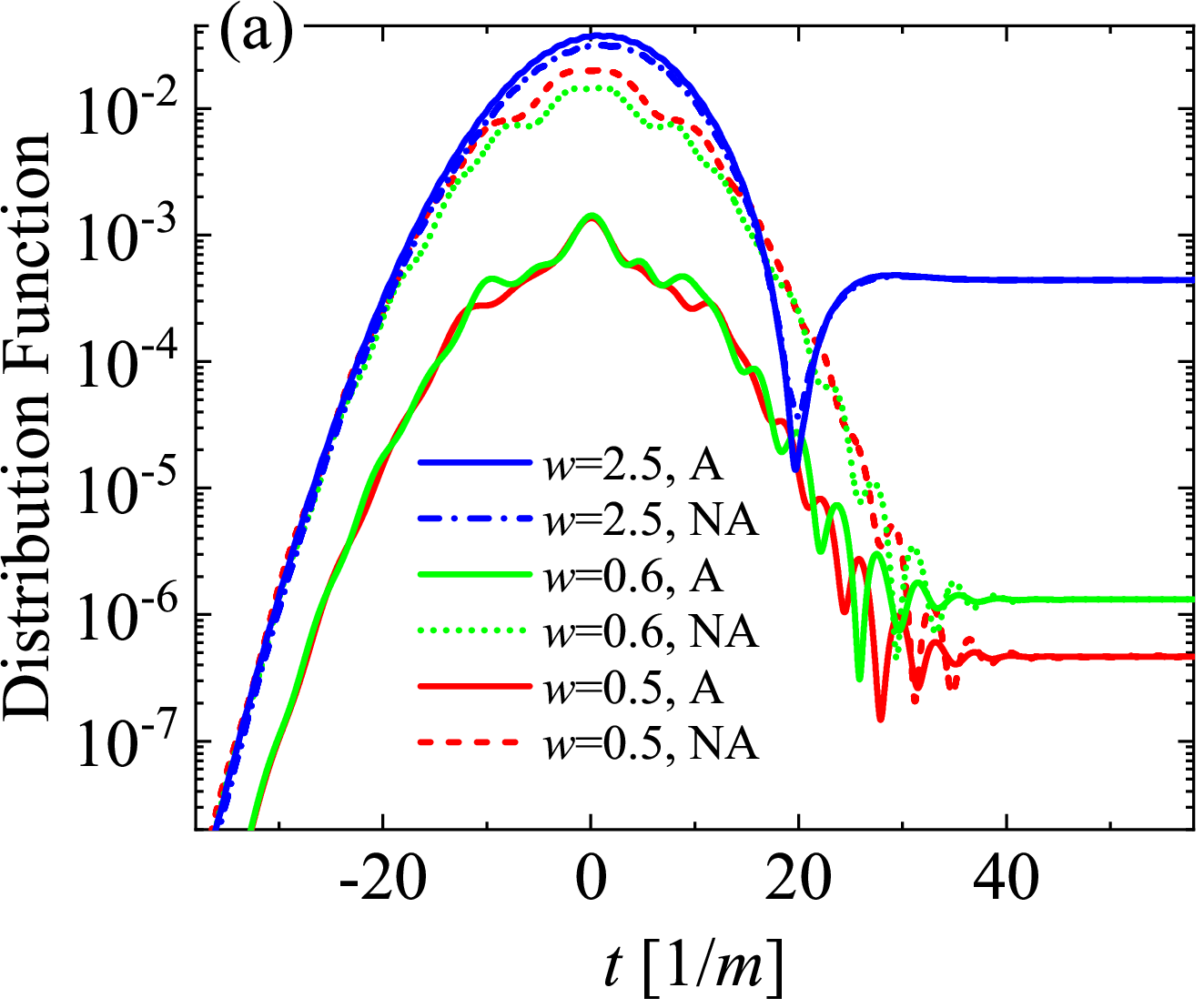}%
\\
\includegraphics[width=0.413\textwidth]{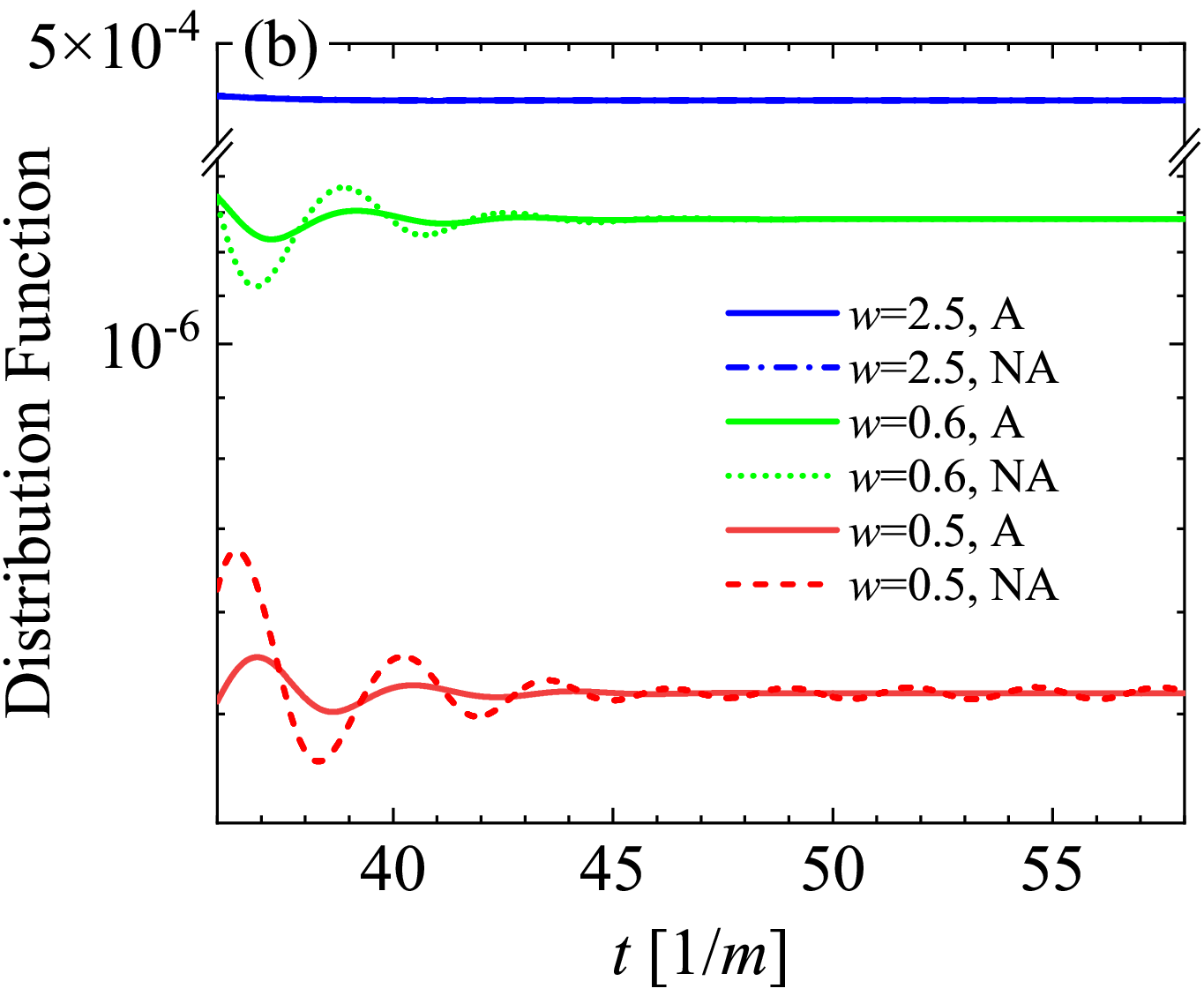}%
\hspace{0.12cm}
\caption{Time evolution of the nonadiabatic (labeled NA) and adiabatic (labeled A) distribution functions of created particles at $k_x=k_z=0,k_y=0.5m$ for a circularly polarized electric field with the frequency $w=0.5m, 0.6m, 2.5m$. (b) is the enlarged drawing of the process of quasiparticles becoming real particles shown in (a). The results are denoted by different lines, see the legend. Other field parameters are $E_0=0.1\sqrt{2}E_{\rm{cr}}$ and $\tau=10/m$.
\label{fig:Fig1}}
\end{figure}

The spatially homogeneous and time-dependent electric field with arbitrary polarization is
\begin{equation}\label{eqn:ElectricField}
\mathbf{E}(t)=\frac{E_0}{\sqrt{1+\delta^2}}\mathrm{exp}
\left(-\frac{t^2}{2\tau^2}\right)\left[\begin{array}{c}\cos(w t+\phi)\\\delta\sin(w t+\phi)\\0\end{array}\right],
\end{equation}
where $E_0/\sqrt{1+\delta^2}$ is the maximum amplitude of the field, $\delta$ represents the polarization, $w$ is the field frequency, $\tau$ denotes the pulse duration, $\sigma=w\tau$ depicts the number of oscillation cycles within the pulse envelope, and $\phi$ is the carrier envelope phase. In our numerical calculations, we fix $\delta=1$ and $\phi=0$. The vector potential of the field is
\begin{equation}\label{eqn:VectorPotential}
\mathbf{A}(t)=\frac{\sqrt{\pi}}{4}e^{-\sigma^2/2}E_0\tau
\left[\begin{array}{c}i\,\mathrm{erfi}(\frac{\sigma}{\sqrt{2}}
+i\frac{t}{\sqrt{2}\tau})\\\mathrm{erfi}(\frac{\sigma}{\sqrt{2}}
-i\frac{t}{\sqrt{2}\tau})\\0\end{array}\right]+\mathrm{c.c.},
\end{equation}
where $\mathrm{erfi}(\cdot)$ is the imaginary error function. It can be seen that with the increase of $\sigma$ the vector potential $\mathbf{A}(\pm\infty)$ trends to zero. This indicates that for a given $E_0$ a large enough $\sigma$ can make $\mathbf{A}(t_f)=\mathbf{A}(t_0)=0$ hold and the nonadiabatic and adiabatic distribution functions
equal at time $t_f$. For example, when $\sigma=5$ in the case of $E_0=0.1\sqrt{2}E_{\mathrm{cr}}$ and $\tau=10/m$, the nonadiabatic and adiabatic distribution functions at $t\rightarrow+\infty$ are different and the former oscillates with time, see Fig. \ref{fig:Fig1}. However, when $\sigma\geq6$ the two distribution functions are almost the same. Therefore, the nonadiabatic and adiabatic distribution functions at the time after the field vanishes are equal for a Gaussian envelope pulse with enough number of oscillation cycles in the envelope.

From Fig. \ref{fig:Fig1}, we also see that for a relatively small field frequency (but $\sigma\geq6$) the derivation of the maximum value of the nonadiabatic distribution function from its stable value is larger than that in the adiabatic case. This indicates that the adiabatic QKEs may provide a better understanding of pair production at intermediate times for a slowly varying electric field. However, for a very large field frequency, for example $w=2.5m$, this situation is exactly the opposite. The nonadiabatic QKEs seem more suitable for depicting pair production at intermediate times than the adiabatic one. Consequently, the nonadiabatic QKEs may be a better choice to describe pair production at intermediate times for a rapidly oscillating electric field.

\begin{figure*}[!ht]
\centering
\includegraphics[width=\textwidth]{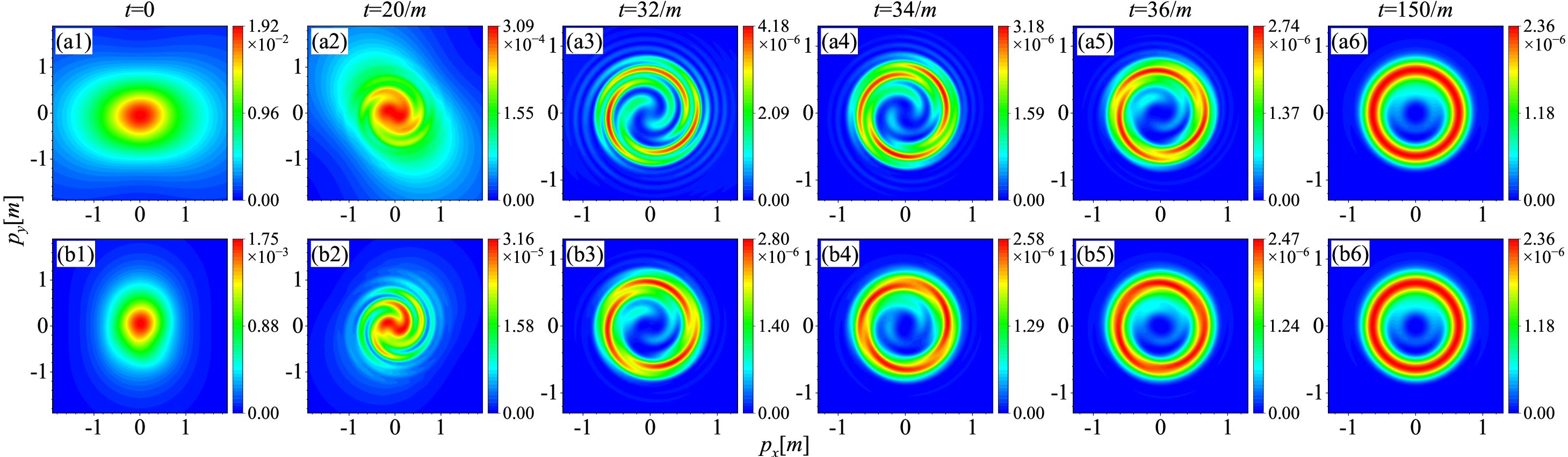}%
\caption{Time evolution of the nonadiabatic (upper row) and adiabatic (lower row) momentum distributions of created particles in the polarized plane ($p_x$, $p_y$) with $p_z=0$ for a circularly polarized electric field. The parameters of the field are $E_0=0.1\sqrt{2}E_{\rm{cr}}$, $w=0.6m$, and $\tau=10/m$.
\label{fig:Fig2}}
\end{figure*}

Figure \ref{fig:Fig2} shows the time evolution of the nonadiabatic (upper row) and adiabatic (lower row) momentum distributions in the polarized plane ($p_x$, $p_y$) at $p_z=0$ for a circularly polarized electric field. One can see that although the nonadiabatic distribution function at intermediate times is generally larger than the adiabatic one, the nonadiabatic momentum distribution has the similar time evolution feature with the adiabatic one, for instance, both of them exhibit the spiral structures which are related to the multi-photon pair production. Furthermore, the number of spirals (three) is exactly one less than the number of photons absorbed in pair production (four). With the disappearance of the electric field the spiral structures vanish and finally form a ring of four-photon absorption. In particular, it is found that Fig. \ref{fig:Fig2} (a5) is almost the same as Fig. \ref{fig:Fig2} (b3).

\section{SUMMARY and DISCUSSION}
\label{sec:five}

In summary, we have derived the nonadiabatic quantum kinetic equations and nonadiabatic DHW formalism for pair production in multi-component electric fields, proved that the two methods are equivalent, and established the relation between the NAQKEs and QKEs. Similarly, the nonadiabatic FVHW formalism is also derived and demonstrated to be equivalent to the NAQVEs in scalar QED.

We have also calculated the time evolution of the nonadiabatic and adiabatic distribution functions of created particles in a circularly polarized Gaussian pulse electric field with a subcycle structure. It is found that if the field has a sufficient number of oscillation cycles, the nonadiabatic and adiabatic distribution functions become the same after the field fades away. However, during the presence of the field, the nonadiabatic distribution function is different from the adiabatic one. Nevertheless, the time evolution signatures of the nonadiabatic momentum distribution are similar to the adiabatic one. For instance, we find that both momentum distributions have spiral structures, and the number of spirals is precisely one less than the number of photons absorbed in multiphoton pair production. This result provides us a more convenient way to determine the number of photons absorbed in multiphoton pair production. Moreover, we also find that for a rapidly oscillating electric fields, the maximum value of the nonadiabatic distribution function is smaller than that of the adiabatic one, which indicates that, to some extent, the NAQKEs and NADHW formalism may give a more meaningful description of pair production at intermediate times.

The derivation of NAQKEs and NADHW formalism will extend the research scope of the nonadiabatic pair production. For example, one can use them to explore the nonadiabatic pair production in arbitrarily polarized electric fields, the spiral structures in nonadiabatic momentum distributions, and the helicity signatures in nonadiabatic pair production. Additionally, since the number density of particles created from the field-free vacuum at time $t$ where the vector potential is different from the initial one is divergent, its renormalization is also worth further study.

\begin{acknowledgments}
The work is supported by the National Natural Science Foundation of China (NSFC) under Grants No. 11974419 and No. 11705278, the Strategic Priority Research Program of Chinese Academy of Sciences (Grant No. XDA25051000, XDA25010100), and the Fundamental Research Funds for the Central Universities (No. 2023ZKPYL02).
\end{acknowledgments}

\appendix
\section{Derivation of the nonadiabatic FVHW formalism}
\label{appa}

In this appendix, we will derive the nonadiabatic Feshbach-Villars-Heisenberg-Wigner (NAFVHW) formalism and prove that it is equivalent to the NAQVE in scalar QED shown in Ref. \cite{LiPRD2024}.

According to Ref. \cite{LiPRD2019}, we know that for a spatially homogeneous and time-dependent electric field the FVHW formalism is
\begin{eqnarray}
\dot{\mathbbm{x}}_0&=&\frac{\mathbf{p}^2}{m}\mathbbm{x}_2,
\label{eqn:FVHW-1}\\
\dot{\mathbbm{x}}_1&=&-\Big(\frac{\mathbf{p}^2}{m}+2m\Big)
\mathbbm{x}_2, \label{eqn:FVHW-2}\\
\dot{\mathbbm{x}}_2&=&\frac{\mathbf{p}^{2}}{m}\mathbbm{x}_0
+\Big(\frac{\mathbf{p}^{2}}{m}+2m\Big)\mathbbm{x}_1,
\label{eqn:FVHW-3}
\end{eqnarray}
with the initial conditions: $\mathbbm{x}_0(t_0)=\frac{\mathbf{p}^2_0+2m^2}{2m\omega_0}, \, \mathbbm{x}_1(t_0)=-\frac{\mathbf{p}_0^2}{2m\omega_0}, \, \mathbbm{x}_2(t_0)=0$,
and the distribution function of created particles from the adiabatic vacuum in the electric field,
\begin{equation}\label{eqn:FVHW-DF}
\mathbbm{f}=\frac{\big(\frac{\mathbf{p}^2}{2m}+m\big)\mathbbm{x}_0
+\frac{\mathbf{p}^2}{2m}\mathbbm{x}_1}{2\omega}-\frac{1}{2}.
\end{equation}

Similar to the NADHW formalism, the NAFVHW formalism is obtained by defining the distribution function as
\begin{equation}\label{eqn:NAFVHW-DF}
\mathfrak{f}=\frac{\big(\frac{\mathbf{p}^2_0}{2m}+m\big)\mathbbm{x}_0
+\frac{\mathbf{p}^2_0}{2m}\mathbbm{x}_1}{2\omega_0}-\frac{1}{2}.
\end{equation}
This distribution function describes the particles created from the field-free vacuum in the electric field. The transformation from FVHW components to NAQVE in scalar QED [cf. Eqs. (A15) in Ref. \cite{LiPRD2024}] is
\begin{eqnarray}
\mathbbm{x}_0&=&\frac{1}{2m\omega_0}[(\mathbf{p}^2_0+2m^2)
(1+2f^-)-\mathbf{p}^2_0h^-],\label{eqn:TF-FVHW-1}\\
\mathbbm{x}_1&=&\frac{1}{2m\omega_0}[-\mathbf{p}^2_0(1+2f^-)+
(\mathbf{p}^2_0+2m^2)h^-],\label {eqn:TF-FVHW-2}\\
\mathbbm{x}_2&=&-g^-.\label{eqn:TF-FVHW-3}
\end{eqnarray}
The inverse transformation is
\begin{eqnarray}
f^-&=&\frac{1}{4m\omega_0}[(\mathbf{p}^2_0+2m^2)\mathbbm{x}_0
+\mathbf{p}^2_0\mathbbm{x}_1]-\frac{1}{2},\label{eqn:TFI-FVHW-1}\\
g^-&=-&\mathbbm{x}_2,\label{eqn:TFI-FVHW-2}\\
h^-&=&\frac{1}{2m\omega_0}[\mathbf{p}^2_0\mathbbm{x}_0
+(\mathbf{p}^2_0+2m^2)\mathbbm{x}_1].\label{eqn:TFI-FVHW-3}
\end{eqnarray}

Moreover, by setting $\mathbbm{x}_0=\frac{\mathbf{p}^2_0+2m}{m\omega_0}(\mathfrak{f}
+\frac{1}{2})-\frac{\mathbf{p}^2_0}{m\omega_0}\mathfrak{x}$ and $\mathbbm{x}_1=\frac{\mathbf{p}^2_0+2m}{m\omega_0}\mathfrak{x}-\frac{\mathbf{p}^2_0}{m\omega_0}(\mathfrak{f}
+\frac{1}{2})$, we can get another version of the NAFVHW formalism:
\begin{eqnarray}\label{eqn:NAFVHW-Formalism}
\dot{\mathfrak{f}}&=&\frac{\omega^2-\omega^2_0}{2\omega_0}
\mathbbm{x}_2, \\
\dot{\mathbbm{x}}_2&=&\frac{2(\omega^2-\omega^2_0)}{\omega_0}
(\mathfrak{f}+\frac{1}{2})+\frac{2(\omega^2+\omega^2_0)}{\omega_0}
\mathfrak{x}, \\
\dot{\mathfrak{x}}&=&-\frac{\omega^2+\omega^2_0}{2\omega_0}\mathbbm{x}_2.
\end{eqnarray}
The initial conditions are $\mathfrak{f}(t_0)=\mathbbm{x}_2(t_0)=\mathfrak{x}(t_0)=0$.
Furthermore, with the replacement, $\mathbbm{x}_2\rightarrow-\mathfrak{g}$ and $\mathfrak{x}\rightarrow\mathfrak{h}/2$, we have
\begin{eqnarray}\label{eqn:NAQVE-Scalar}
\dot{\mathfrak{f}}&=&\frac{\omega^2_0-\omega^2}{2\omega_0}
\mathfrak{g}, \\
\dot{\mathfrak{g}}&=&\frac{\omega^2_0-\omega^2}{\omega_0}
(1+2\mathfrak{f})-\frac{\omega^2_0+\omega^2}{\omega_0}
\mathfrak{h}, \\
\dot{\mathfrak{h}}&=&\frac{\omega^2_0+\omega^2}{\omega_0}\mathfrak{g},
\end{eqnarray}
with the initial conditions $\mathfrak{f}(t_0)=\mathfrak{g}(t_0)=\mathfrak{h}(t_0)=0$. One can see that the above equations are just the NAQVE in scalar QED given in Eqs. (A15) in Ref. \cite{LiPRD2024}.

\end{document}